\documentclass[10pt]{revtex4}
\usepackage{amssymb}
\usepackage{latexsym}
\usepackage{epsfig}
\begin{document}
\title{Higher dimensional dilaton black holes in the presence of exponential nonlinear electrodynamics }
\author{A. Sheykhi$^{1,2}$\footnote{asheykhi@shirazu.ac.ir} and  A. Kazemi $^{1}$}
\address{$^1$ Physics Department and Biruni Observatory, College of
Sciences, Shiraz University, Shiraz 71454, Iran\\
$^2$ Research Institute for Astronomy and Astrophysics of Maragha
(RIAAM), P.O. Box 55134-441, Maragha, Iran}
\begin{abstract}
We examine the higher dimensional action in which gravity is
coupled to the exponential nonlinear electrodynamic and a scalar
dilaton field. We construct a new class of $n$-dimensional static
and spherically symmetric black hole solutions of this theory in
the presence of the dilaton potential with two Liouville-type
terms. In the presence of two Liouville-type dilaton potential,
the asymptotic behavior of the obtained black holes are neither
flat nor (A)dS. Due to the nonlinear nature of electrodynamic
field, the electric field has finite value near the origin where
$r\rightarrow0$ and goes to zero as $r\rightarrow\infty$.
Interestingly enough, we find that in the absence of the dilaton
field, the electric field has a finite value at $r=0$, while as
soon as the dilaton field is taken into account, the electric
field diverges as $r\rightarrow 0$. This implies that the presence
of the dilaton field changes the behaviour of the electric field
near the origin. In the limiting case where the nonlinear
parameter $\beta$ goes to infinity, our solutions reduce to
dilaton black holes of Einstein-Maxwell-dilaton gravity in higher
dimensions. We compute the conserved and thermodynamic quantities
of the solutions and show that these quantities satisfy the first
law of black holes
thermodynamics on the horizon. 
\end{abstract}
\maketitle

\section{Introduction}
There are several motivations for studying black hole solutions in
higher dimensional spacetimes. The first reason comes from string
theory which contains gravity and requires more than four
dimensions. In fact, the first successful statistical counting of
black hole entropy in string theory was performed for a five
dimensional black hole \cite{Stro}. This example provides the best
laboratory for the microscopic string theory of black holes.
Another motivation originates from the AdS/CFT correspondence
which relates the properties of an $n$-dimensional black hole with
those of a quantum field theory in  $(n-1)$-dimensions \cite{Ah}.
Besides, the production of higher-dimensional black holes in
future colliders becomes a conceivable possibility in scenarios
involving large extra dimensions and TeV-scale gravity \cite{MC}.
In addition, as mathematical objects, black hole spacetimes are
among the most important Lorentzian Ricci-flat manifolds in any
dimension \cite{Emp}.

In this paper, we turn to the investigation on higher dimensional
black holes in the presence of nonlinear electrodynamics and
dilaton field. The theory of nonlinear electrodynamics was first
introduced in 1930's by Born and Infeld to obtain a classical
theory of charged particles with finite self-energy \cite{BI}.
Born-Infeld (BI) theory has received renewed attentions since it
turns out to play an important role in string theory. It arises
naturally in open superstrings and D-branes \cite{Frad,Cal}. The
low energy effective action for an open superstring in loop
calculations lead to BI type actions. The BI action including a
dilaton and an axion field, appears in the coupling of an open
superstring and an Abelian gauge field theory \cite{Frad}. This
action, describing a Born-Infeld-dilaton-axion system coupled to
Einstein gravity, can be considered as a nonlinear extension in
the Abelian field of Einstein-Maxwell-dilaton-axion gravity.
Nonlinear BI theory in the context of dilaton gravity have been
investigated by many authors. See for e.g \cite{Tam1,Tam2,YI,yaz,
Clement,yazad,SRM,Shey1,Shey2,DHSR} and references therein. In the
framework of Einstein-Maxwell-dilaton (EMd) gravity with a
Liouville type dilaton potential non-degenerate and degenerate
(extremal) Killing horizons of arbitrary geometry and topology
were investigated \cite{Shohreh}. Here we would like to consider
another type of nonlinear electrodynamics, namely exponential form
of the nonlinear electrodynamics in the setup of dilaton gravity.
The lagrangian of the exponential nonlinear (EN) electrodynamics
is given by \cite{HendiJHEP},
\begin{equation}\label{LEN}
L_{\rm EN}=4\beta^{2}
\left[\exp\left(-\frac{F^2}{4\beta^2}\right)-1\right],
\end{equation}
where $\beta $ is called the nonlinear parameter with dimension of
mass, $F^2=F_{\mu \nu }F^{\mu \nu }$, where $F_{\mu \nu }$ is the
electromagnetic field tensor. The advantages of the EN Lagrangian
compared to BI nonlinear electrodynamics is that, it does not
cancel, at least for some parameters, the divergency of the
electric field at $r = 0$, however, its singularity is much weaker
than Einstein-Maxwell theory \cite{Hendi2}. In our previous paper
\cite{ShHaj}, we studied dilaton black holes coupled to EN
electrodynamics in four dimensional spacetime. In the present
work, we would like to extend our study to all higher dimensions
and construct a new class of $n$-dimensional black holes in
dilaton gravity which is coupled to nonlinear matter field. We
shall also investigate the physical properties of the spacetime
and obtain conserved and thermodynamic quantities of the
solutions.

This paper is organized as follows. In the next section, we
introduce the Lagrangian of EN electrodynamics in $n$-dimensions
coupled to the dilaton field in Einstein gravity, and obtain the
corresponding field equations by varying the action. In section
\ref{solu}, we find a new class of static and spherically
symmetric black hole solutions of this theory and investigate
their properties. In section \ref{Therm}, we study thermodynamics
of higher dimensional dilaton black holes in the presence of
nonlinear electrodynamics. The last section is devoted to
conclusion and discussion.

\section{Basic field equations}\label{Field}
We consider the $n$-dimensional $(n\geq4)$ action in which gravity
is coupled to dilaton and nonlinear electrodynamic fields
\begin{equation}\label{Act}
S=\frac{1}{16\pi}\int{d^{n}x\sqrt{-g}\left(\mathcal{R}\text{
}-\frac{4}{n-2}(\nabla \Phi )^{2}-V(\Phi )+L(F,\Phi)\right)},
\end{equation}
where $\mathcal{R}$ and  $\Phi $ are, respectively, the Ricci
scalar curvature and the dilaton filed, and $V(\Phi )$ is a
potential for $\Phi $. We further assume the dilaton potential
contain two Liouville terms,
\begin{equation}\label{v2}
V(\Phi) = 2\Lambda_{0} e^{2\zeta_{0}\Phi} +2 \Lambda e^{2\zeta
\Phi},
\end{equation}
where $\Lambda_{0}$,  $\Lambda$, $ \zeta_{0}$ and $ \zeta$ are
constants. This kind of potential was previously investigated in
the context of BI-dilaton (BId) black holes \cite{Shey1,Shey2} as
well as EMd gravity \cite{MW,CHM,Cai,yaz2,Clem,Shey3}.

We choose the Lagrangian of the exponential nonlinear
electrodynamics coupled to the dilaton field (ENd) in
$n$-dimensions as
\begin{equation}\label{LFPhi}
L(F,\Phi)=4\beta^{2} e^{4\alpha
\Phi/(n-2)}\left[\exp\left(-\frac{e^{- 8\alpha
\Phi/(n-2)}F^2}{4\beta^{2}}\right)-1\right],
\end{equation}
where $\alpha$ is a constant determining the strength of coupling
of the scalar and electromagnetic field. In order to justify such
a choice for the Lagrangian of ENd field, let us invoke the BId
Lagrangian in $n$-dimensions which is written as \cite{Shey1}
\begin{equation}\label{LFPhiB}
L_{\rm BId}(F,\Phi)=4\beta ^{2}e^{4\alpha \Phi /(n-2)}\left( 1-\sqrt{1+\frac{%
e^{-8\alpha \Phi /(n-2)}F^{2}}{2\beta ^{2}}}\right).
\end{equation}
It is worthy to note that Lagrangian (\ref{LFPhiB}) originates
from open string version of the BI action coupled to a dilaton
field and only valid for the pure electric case \cite{Tam1}.
Clearly, this version of the BId action does not enjoy
electric-magnetic duality \cite{Clement}. This form for the BId
term have been investigated previously by a number of authors
\cite{Tam1,Tam2,YI,yaz, Clement,yazad,SRM,Shey1,Shey2,DHSR}. It is
easy to check that the series expansion of both BId Lagrangian
(\ref{LFPhiB}) and ENd Lagrangian (\ref{LFPhi}), for large
$\beta$, have the same behavior
\begin{equation}\label{ExB}
L_{\rm BId}(F,\Phi)= -e^{-4\alpha \Phi/(n-2)}F^2+\frac{
e^{-12\alpha \Phi/(n-2)}F^4}{8\beta^{2}}-\frac{ e^{-20\alpha
\Phi/(n-2)}F^6}{32\beta^{4}}+O\left(\frac{1}{\beta^6}\right),
\end{equation}
\begin{equation}\label{ExE}
L(F,\Phi)=-e^{-4\alpha \Phi/(n-2)}F^2+\frac{ e^{-12\alpha
\Phi/(n-2)}F^4}{8\beta^{2}}-\frac{ e^{-20\alpha
\Phi/(n-2)}F^6}{96\beta^{4}}+O\left(\frac{1}{\beta^6}\right).
\end{equation}
This similarity implies that one can either consider (\ref{LFPhi})
or (\ref{LFPhiB}) as the nonlinear electrodynamic Lagrangian
coupled to the dilaton field. Here we would like to study the new
ENd Lagrangian (\ref{LFPhi}), and investigate the effects of this
kind of nonlinear electrodynamics coupled to the dilaton field on
the behavior of the solutions. In the absence of the dilaton field
($\alpha=0$) and in four dimensions where $n=4$, $L(F,\Phi)$
reduces to EN electrodynamic Lagrangian presented in
\cite{Hendi2}. On the other hand, in the limiting case where
$\beta\rightarrow\infty $, both $L(F,\Phi)$ and $L_{\rm
BId}(F,\Phi)$ recovers the standard linear Maxwell lagrangian
coupled to the dilaton field in $n$-dimensions \cite{CHM}
\begin{equation}
L(F,\Phi)=L_{\rm BId}(F,\Phi)=-e^{-4\alpha \Phi/(n-2)}F^2.
\end{equation}
This is an expected result, since in this case the nonlinear
electrodynamics reduces to the linear Maxwell electrodynamics. For
latter convenience we rewrite
\begin{equation}
L(F,\Phi)=4\beta^{2} e^{4\alpha \Phi/(n-2)}{\mathcal{L}}(Y),
\end{equation}
where
\begin{eqnarray}
{\mathcal{L}}(Y) &=&\exp(-Y)-1,\label{LY}\\
Y&=& \frac{e^{-8\alpha \Phi/(n-2)}F^2}{4\beta^2}.\label{Y}
\end{eqnarray}
By varying action (\ref{Act}) with respect to the gravitational
field $g_{\mu \nu }$, the dilaton field $\Phi $ and the gauge
field $A_{\mu }$ we obtain the field equations as
\begin{eqnarray}\label{FE1}
{\cal R}_{\mu\nu}&=& \frac{4}{n-2} \left(\partial _{\mu }\Phi
\partial _{\nu }\Phi +\frac{1}{4}g_{\mu \nu }V(\Phi)\right)-
2e^{-4\alpha \Phi/(n-2)}\partial_{Y}{{\cal L}}(Y) F_{\mu\eta}
F_{\nu}^{\text{ }\eta } \nonumber \\
&&+\frac{4\beta^2}{n-2} e^{4\alpha \Phi/(n-2)}
\left[2Y\partial_{Y}{{\cal L}}(Y)-{{\cal L}}(Y)\right]g_{\mu\nu},
\end{eqnarray}
\begin{equation}\label{FE2}
\nabla ^{2}\Phi =\frac{n-2}{8}\frac{\partial V}{\partial \Phi}+
2\alpha \beta^2 e^{4\alpha \Phi/(n-2) }\left[2{
Y}\partial_{Y}{{\cal L}}(Y)-{\cal L}(Y)\right],
\end{equation}
\begin{equation}\label{FE3}
\nabla _{\mu }\left(e^{-4\alpha \Phi/(n-2)}
\partial_{Y}{{\cal L}}(Y) F^{\mu\nu}\right)=0.
\end{equation}
When $\beta\rightarrow\infty$, we have ${\cal L}(Y)=-Y$, and the
system of field equations (\ref{FE1})-(\ref{FE3}) restore the
well-known equations of EMd gravity
\cite{CHM,Cai,yaz2,Clem,Shey3}.
\section{higher dimensional dilaton black hole}\label{solu}
In this section, we would like to find static and spherically
symmetric solutions of the field equations
(\ref{FE1})-(\ref{FE3}). The metric of such a spacetime can be
written
\begin{equation}\label{metric}
ds^2=-f(r)dt^2 + {dr^2\over f(r)}+ r^2R^2(r)d\Omega^2_{n-2},
\end{equation}
where $d\Omega^2_{n-2}$ denotes the metric of an unit
$(n-2)$-sphere and $f(r)$ and $R(r)$ are functions of $r$ which
should be determined. Integrating the electromagnetic field
equation (\ref{FE3}) and assuming all components of $F_{\mu\nu}$
are zero except $ F_{tr}$, we find
\begin{equation}\label{FtrE}
F_{tr}=\frac{q e^{4\alpha \Phi /(n-2)}}{\left( rR\right) ^{n-2}}
\exp\left[-\frac{1}{2}L_W\left(\frac{q^2}{\beta^2 \left( rR\right)
^{2n-4}}\right)\right],
\end{equation}
where $q$ is an integration constant which is related to the
electric charge of the black hole. We recall that in BId theory
the electric field can be written as \cite{Shey1}
\begin{equation}\label{Ftr}
F_{tr}=\frac{q e^{4\alpha \Phi/(n-2)}}{\left(
rR\right)^{n-2}}\sqrt{1+\frac{2q^2}{\beta^2 (n-2) \left(
rR\right)^{2n-4}}}.
\end{equation}
Using the Gauss's law, $ Q = \frac{1}{4\pi} \int
\exp\left[{-4\alpha\Phi/(n-2)}\right]  \text{ }^{*} F d{\Omega}, $
we can calculate the flux of the electromagnetic field at infinity
to obtain the electric charge of black hole as
\begin{equation}
{Q}=\frac{q\omega _{n-2}}{4\pi},  \label{Charge}
\end{equation}
where $\omega_{n-2}$ represents the volume of an unit
$(n-2)$-sphere. Note that in (\ref{FtrE}), $L_W(x)={LambertW(x)}$
is the Lambert function which satisfies the identity
\cite{Lambert}
\begin{equation}
L_W(x)e^{L_W(x)}=x,
\end{equation}
and has the following series expansion
\begin{equation}\label{LW}
L_W(x)=x-x^2+\frac{3}{2}x^3-\frac{8}{3}x^4+....
\end{equation}
Clearly, series (\ref{LW}) converges for $|x| <1$. In the absence
of the nonlinear dilaton field ($\alpha=0$) we have $R(r)=1$, and
Eq. (\ref{FtrE}) reduces to
\begin{equation}\label{FtrE2}
F_{tr}=\frac{q }{r^{n-2}}
\exp\left[-\frac{1}{2}L_W\left(\frac{q^2}{\beta^2 r^{2n-4}
}\right)\right],
\end{equation}
while in the limiting case where $\beta \rightarrow \infty $, it
reduces to the electric field of $n$-dimensional EMd black holes
\cite{Shey3}
\begin{equation}
F_{tr}=\frac{q e^{4\alpha \Phi/(n-2)}}{(
rR)^{n-2}}+O\left(\frac{1}{\beta^{2}}\right). \label{FtrM}
\end{equation}
In order to solve the system of equations (\ref{FE1}) and
(\ref{FE2}) for three unknown functions $f(r)$, $R(r)$ and $\Phi
(r)$, we make the ansatz \cite{Shey3}
\begin{equation} \label{Rphi}
R(r)=e^{2\alpha \Phi /(n-2)}.
\end{equation}
It is important to note that in the absence of the nontrivial
dilaton field ($\alpha=0$), we have $R(r)=1$, as one expected (see
Eq. \ref{metric}). Substituting (\ref{Rphi}), the electromagnetic
field (\ref{FtrE}) and the metric (\ref{metric}) into the field
equations (\ref{FE1}) and (\ref{FE2}), one can obtain the
following solutions
\begin{eqnarray}
f(r)&=&-\frac{{(n-3)}{
{({\alpha}^2+1)}^{2}}}{{({\alpha}^2+n-3)}{({\alpha}^{2}-1)}
}{b}^{-\gamma}{r}^{\gamma}-\frac{m}{r^{n-3-(n-2)\gamma/2}}+\frac{2{{(\alpha^2+1)}^{2}}{(\Lambda+2\beta^2)}{{b}^{\gamma}}}{{(n-2)}{({\alpha}^{2}-n+1)}
}r^{2-\gamma}\nonumber \\
&&-\frac{4\beta q \left( \alpha ^{2}+1\right)
b^{{(4-n)\gamma}/{2}}} {{(n-2)}{{r}^{n-3-(n-2)\gamma/2}}} \int r
^{-\gamma}\left(\sqrt{L_{W}(\eta)}-\frac{1}{\sqrt{L_{W}(\eta)}}\right){dr},
\label{f1}
\end{eqnarray}
\begin{equation}\label{phi}
\Phi (r)=\frac{(n-2)\alpha }{2(\alpha ^{2}+1)}\ln
\left(\frac{b}{r}\right), \label{phi}
\end{equation}
where $b$ is an arbitrary constant, $\gamma =2\alpha
^{2}/(1+\alpha ^{2})$, and
\begin{equation}
\eta\equiv \frac{q^{2} b^{(2-n)\gamma}}{\beta
^{2}r^{(n-2)(2-\gamma)}}. \label{eta}
\end{equation}
In the above expression, $m$ appears as an integration constant
and is related to the mass of the black hole. The obtained
solutions fully satisfy the system of equations (\ref{FE1}) and
(\ref{FE2}) provided we take
\begin{equation}\label{lam}
\zeta_{0} =\frac{2}{\alpha(n-2)},   \hspace{.8cm}
\zeta=\frac{2\alpha}{n-2}, \hspace{.8cm}    \Lambda_{0} =
\frac{(n-2)(n-3)\alpha^2 }{2b^2(\alpha^2-1)}.
\end{equation}
Notice that here  $\Lambda$ remains as a free parameter which
plays the role of the cosmological constant. For later
convenience, we redefine it as $\Lambda=-(n-1)(n-2)/2l^2$, where
$l$ is a constant with dimension of length. The integration of Eq.
(\ref{f1}) can be performed using the Mathematica software. The
resulting solution can be written
\begin{eqnarray}\label{f2}
f(r)&=&-\frac{{(n-3)}{({\alpha}^{2}+1)^2}}{{({\alpha}^{2}-1)}{({\alpha}^2+n-3)}
}{b}^{-\gamma}{r}^{\gamma}-\frac{m}{r^{n-3-(n-2)\gamma/2}}+\frac{2(\Lambda+2\beta^2)
\left( {\alpha}^{2}+1 \right)
^{2}{b}^{\gamma}}{{(n-2)}{(\alpha^2-n+1)} }r^{2-\gamma}\nonumber
\\&&+\frac{2\beta q (\alpha^2+1)^4 b^{{(4-n)\gamma}/{2}}}{
{(n-2)^2}{(\alpha^2-1)^2}}\left(\frac{\beta^2
b^{(n-2)\gamma}}{q^2}\right)^{\frac{1-\gamma}{(n-2)(\gamma-2)}}r^{{(n-2)\gamma}/{2}-n+3}
\nonumber
\\ &&\times
\left(\frac{1-\alpha^2}{2n-4}\right)^{\frac{2-2n+\gamma
n}{(\gamma-2)(2n-4)}} \Bigg{\{} -(n-2)^2(\gamma-2)^2\left[\Gamma
\left(\frac{\alpha^2+3n-7}{2n-4},\frac{1-\alpha^2}{2n-4}L_{W}(\eta)\right)-\Gamma\left({\frac{\alpha^2+3n-7}{2n-4}}\right)\right]\nonumber \\
&&+(\gamma-1)^2\left[\Gamma
\left(\frac{\alpha^2-n+1}{2n-4},\frac{1-\alpha^2}{2n-4}L_{W}(\eta)\right)
-\Gamma\left({\frac{\alpha^2-n+1}{2n-4}}\right)\right]\Bigg{\}}.
\end{eqnarray}
where $\Gamma(a,z)$ and $\Gamma(a)$ are Gamma functions and they
are related to each other via,
\begin{eqnarray}\label{Gamma}
\Gamma(a,z)=\Gamma(a)-\frac{z^a }{a}\digamma\left(a,1+a,-z\right).
\end{eqnarray}
where $ \digamma(a,b,z)$ is hypergeometric function
\cite{Lambert}. Using (\ref{Gamma}), solution (\ref{f2}) can also
be reexpressed in terms of hypergeometric function,
\begin{eqnarray}\label{fhyper}
f(r)&=&-\frac{(n-3)({\alpha}^{2}+1)^2}{({\alpha}^{2}-1)({\alpha}^{2}+n-3)
}{b}^{-\gamma}{r}^{\gamma}-\frac{m}{r^{n-3-{(n-2)\gamma}/{2}}}+\frac{2(\Lambda+2\beta^2)
\left( {\alpha}^{2}+1 \right)
^{2}{b}^{\gamma}}{(n-2)(\alpha^2+1-n)
}r^{2-\gamma}\nonumber\\&&+\frac{4 \beta
q{(\alpha^2+1)^2}b^{(4-n)\gamma/2}}{(n-2)r^{n-3-(n-2)\gamma/2}}
{\left(\frac{\beta^2
b^{(n-2)\gamma}}{q^2}\right)}^{\frac{1-\gamma}{(n-2)(\gamma-2)}}{{
L_W^{\frac{\alpha^2-n+1}{2n-4}}}(\eta)} \nonumber
\\ &&\times\Bigg{\{}\frac{L_W^2(\eta)}{\alpha^2+3n-7}\digamma\left(\frac{\alpha^2+3n-7}{2n-4},\frac{\alpha^2+5n-11}{2n-4},\frac{\alpha^2-1}{2n-4}L_W(\eta)\right)\nonumber\\&&
-\frac{1}{\alpha^2-n+1}\digamma\left(\frac{\alpha^2-n+1}{2n-4},\frac{\alpha^2+n-3}{2n-4},\frac{\alpha^2-1}{2n-4}L_W(\eta)\right)
\Bigg{\}}.
\end{eqnarray}
Using the fact that $L_W(x)$ has a convergent series expansion for
$|x| <1$ as given in (\ref{LW}), we can expand (\ref{f2}) for
large $\beta$. The result is
\begin{eqnarray}\label{f3}
f(r)&=&-\frac{(n-3)(\alpha^2+1)^2}{(\alpha^2+n-3)(\alpha^2-1)}b^{-\gamma}
r^\gamma-\frac{m}{r^{n-3-{(n-2)\gamma}/{2}}}+\frac{2\Lambda(\alpha^2+1)^2}{(n-2)(\alpha^2-n+1)}
b^\gamma r^{2-\gamma}\nonumber\\
&&+\frac{2
q^2({\alpha^2+1})^2b^{-(n-3)\gamma}}{(n-2)(\alpha^2+n-3)r^{-(n-3)\gamma+2n-6}}-\frac{q^4
(\alpha^2+1)^2b^{-(2n-5)\gamma}}{2\beta^2(n-2)(\alpha^2+3n-7)r^{(2n-5)(2-\gamma)}}+O\left(\frac{1}{\beta^4}\right).
\end{eqnarray}
When $\beta\rightarrow\infty$, solution (\ref{f3}) is exactly the
one obtained for higher dimensional black holes in EMd gravity
\cite{Shey3}. This is an expected result, since as we discussed
already in this limit the Lagrangian of ENd theory reduces to EMd
gravity. In the absence of the dilaton field ($\alpha =0=\gamma$),
solution (\ref{f3}) can be further simplified as
\begin{eqnarray}\label{f4}
f(r) &=&1-\frac{m}{r^{n-3}}+\frac{r^2}{l^2}
+\frac{2q^{2}}{{(n-2)(n-3)r^{2n-6}}}- \frac{1}{2
\beta^2(n-2)(3n-7)}\frac{q^{4}}{r^{4n-10}}+O\left(\frac{1}{\beta^4}\right),
\end{eqnarray}
which has the form of static spherically symmetric $n$-dimensional
RN-AdS black holes in the limit $\beta\rightarrow \infty$. The
last term in the right hand side of (\ref{f4}) is the leading
nonlinear correction term to the RN-AdS black hole in the large
$\beta$ limit.

\subsection*{Physical properties of the solutions} Now we back to the electric field
obtained in (\ref{FtrE}). In order to study the behaviour of the
electric field, we combine Eqs. (\ref{Rphi}) and (\ref{phi}) with
(\ref{FtrE}). We find
\begin{equation}\label{FtrE2}
F_{tr}=E(r)=\frac{q b^{(4-n)\gamma/2}}{r^{n-2+2\gamma-n \gamma/2}}
\exp\left[-\frac{1}{2}L_W\left(\frac{q^{2} b^{-(n-2)\gamma}}{\beta
^{2} r^{(n-2)(2-\gamma)}}\right)\right].
\end{equation}
{Expanding for large $\beta$, we arrive at}
\begin{equation}\label{FtrE3}
E(r)=\frac{q b^{(4-n)\gamma/2}}{r^{n-2+(4-n)
\gamma/2}}-\frac{q^3}{2\beta^2}\frac{b^{\gamma(4-3n/2)}}{r^{3n-6+\gamma(4-3n/2)}}+
\frac{5q^5}{8\beta^4}\frac{b^{\gamma(6-5n/2)}}{r^{5n-10+\gamma(6-5n/2)}}+O\left(\frac{1}{\beta^6}\right).
\end{equation}
{In order to analyze the behaviour of $F_{tr}$, we choose $b=1$,
$n=6$ and $\alpha=\sqrt{2}$ ($\gamma=4/3$). For these values of
the parameters we have}
\begin{equation}\label{FtrE4}
E(r)=\frac{q}{r^{8/3}}-\frac{q^3}{2\beta^2 r^{16/3}}+
\frac{5q^5}{8\beta^4 r^8}+O\left(\frac{1}{\beta^6}\right).
\end{equation}
{From Eq. (\ref{FtrE4}) we see that in the presence of the dilaton
field, the electric field diverges as $r\rightarrow0$. For more
details see table A. }
\begin{figure}[tbp]
\epsfxsize=7cm \centerline{\epsffile{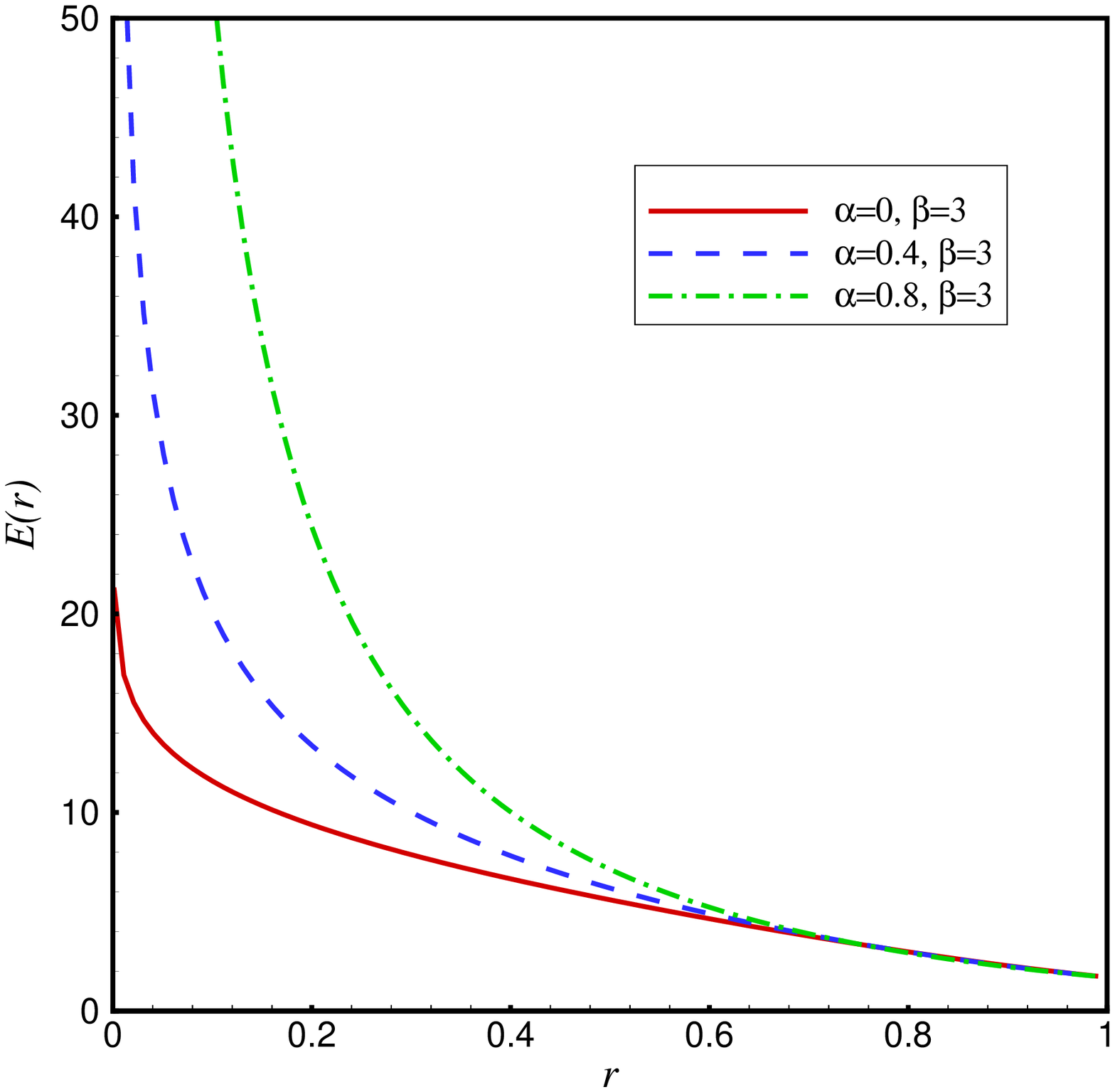}} \caption{The
behavior of the electric field $E(r)$ of ENd black holes versus
$r$ for $n=6$, $b=1$ and $q=2$.} \label{figure1}
\end{figure}
\begin{center}
\begin{tabular}{cccccccc}
\hline\hline $r=$   & $10^{3}$ & $10^{2}$ & $10^{1}$ & $10^{-1}$ &
$10^{-10}$ &  $10^{-1000}$ \\ \hline $\alpha=0.0$ & $10^{-11}$ &
$10^{-7}$ & $10^{-4}$ & $10$ &
$40$ & $400$ \\
$\alpha=0.4$ & $10^{-10}$ & $10^{-7}$ & $10^{-4}$ & $20$ &
$10^4$ & $10^{300}$  \\
$\alpha=0.8$ & $10^{-9}$ & $10^{-6}$ & $10^{-3}$ & $50$ &
$10^{10}$ & $10^{800}$   \\ \hline\hline
\end{tabular}
\\[0pt]
Table A: $E(r)$ for ENd  versus $r$ for $\beta =3$, $n=6$, $q=2$
and $b=1$ and different values of $\alpha$.
\end{center}
\begin{figure}[tbp]
\epsfxsize=7cm \centerline{\epsffile{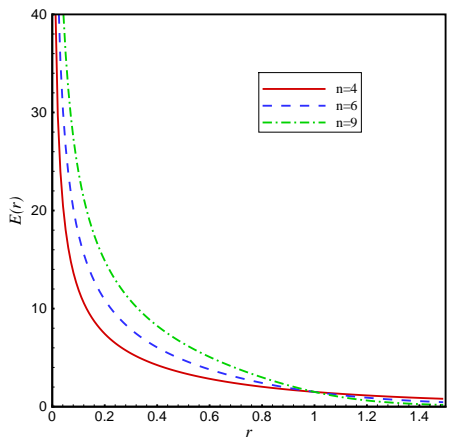}} \caption{The
behavior of the electric field $E(r)$ of ENd black holes versus
$r$ for $b=1$, $\alpha=0.5$, $\beta=2$ and $q=2$.} \label{figure2}
\end{figure}
\begin{figure}[tbp]
\epsfxsize=7cm \centerline{\epsffile{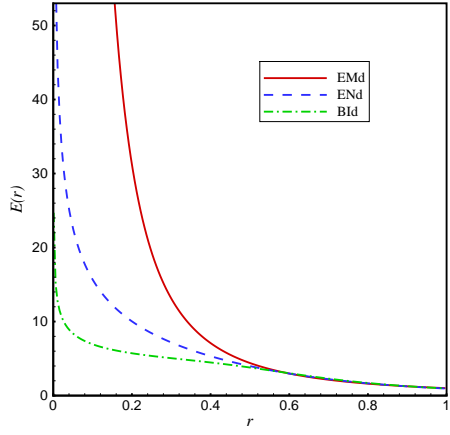}} \caption{The
behavior of the electric field $E(r)$ versus $r$ and  for EMd, ENd
and BId field. Here we have taken $b=1$, $\beta=3$,
$\protect\alpha=0.4$, $n=5$ and $q=1$.} \label{figure3}
\end{figure}
\begin{figure}[tbp]
\epsfxsize=7cm \centerline{\epsffile{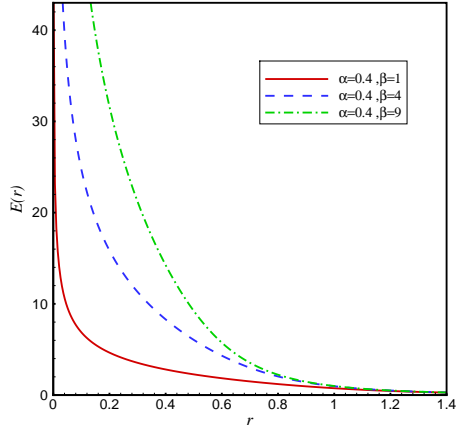}} \caption{The
behavior of the electric field $E(r)$ of ENd black holes versus
$r$ for $b=1$, $n=6$, and $q=1$.} \label{figure4}
\end{figure}
{In order to have better understanding of the behavior of the
electric field (\ref{FtrE2}), we plot $E(r)$ versus $r$ for
different values of the parameters in figures
\ref{figure1}-\ref{figure4}. From these figures and table A we see
that the electric field goes to zero for large $r$ independent of
the value of the other parameters.} Figure \ref{figure1} shows
that for ENd black holes, and in the absence of the dilaton field
($\alpha=0$), the electric field has a finite value near $r=0$,
while as soon as the dilaton field is taken into account
($\alpha>0$), the electric field diverges as $r\rightarrow 0$.
This implies that the presence of the dilaton field changes the
behavior of the electric field near the origin. The behavior of
the electric fields for ENd black holes in different dimensions
are shown in figure \ref{figure2}. From this figure we see that in
any dimension as $r\rightarrow0$, the electric field goes to
infinity. In figures 3 we have compared the behavior of $E(r)$ for
BId, ENd and EMd black holes. From this figure we see that for ENd
case the electric field has a finite value \emph{near} the origin,
while it diverges exactly at $r=0$, however its singularity is
weaker than EMd. This is in contrast to the BId electrodynamics
which the electric field has finite value at $r=0$. Finally, we
have plotted in figure 4 the electric field of ENd black holes for
different values of the nonlinear parameter $\beta$. From this
figure we see that with increasing $\beta$, the electric field
diverges as $r\rightarrow0$. This is an expected result, since for
large $\beta$ our theory reduces to the well-known EMd gravity
\cite{Shey3}.

The next step is to investigate the casual structure of the
solutions and check whether there is or not the curvature
singularities and horizons. We find that  Kretschmann scalar
$R_{\mu \nu \lambda \kappa }R^{\mu \nu \lambda \kappa }$ diverges
as $r\rightarrow 0$. This implies that our spacetime has an
essential singularity located at $r=0$.

{In order to study the asymptotic behaviour of the solutions, we
expand the metric function $f(r)$ for $r\rightarrow\infty$ limit.
We find
\begin{eqnarray}\label{fasymp}
\lim_{r\longrightarrow \infty}f(r)
=-\frac{(n-3)(\alpha^2+1)^2}{(\alpha^2+n-3)(\alpha^2-1)}b^{-\gamma}
r^\gamma+\frac{2\Lambda(\alpha^2+1)^2}{(n-2)(\alpha^2-n+1)}
b^\gamma r^{2-\gamma}.
\end{eqnarray}
Let us note that in the absence of the dilaton field
($\alpha=0=\gamma$), the metric function becomes
\begin{eqnarray}\label{fasymp2}
\lim_{r\longrightarrow \infty}f(r) =1-\frac{2\Lambda
r^2}{(n-1)(n-2)},
\end{eqnarray}
which describes an asymptotically flat ($\Lambda=0$), AdS
($\Lambda<0$) or dS ($\Lambda>0$) spacetimes. However, as one can
see from Eq. (\ref{fasymp}), in the presence of the dilaton field
the asymptotic behaviour is neither flat nor (A)dS. For example,
taking $\alpha=\sqrt{2}$, $n=6$ and $b=1$, we have
\begin{eqnarray}\label{fasymp3}
\lim_{r\longrightarrow \infty}f(r) =-\frac{27}{5}
r^{4/3}-\frac{3\Lambda}{2} r^{2/3}.
\end{eqnarray}
Clearly, the metric function (\ref{fasymp3}) is neither falt nor
(A)dS. Indeed, it has been shown that no dilaton dS or AdS black
hole solution exists with the presence of only one or two
Liouville-type dilaton potential \cite{MW}. In the presence of one
or two Liouville-type potential, black hole spacetimes which are
neither asymptotically flat nor (A)dS have been explored by many
authors (see e.g. \cite{CHM,Cai,yaz2,Clem,Shey3}). It is important
to note that this asymptotic behaviour is not due to the nonlinear
nature of the electrodynamic field, since as $r\rightarrow\infty$
the effects of the nonlinearity disappear. This is due to the fact
that, $r\rightarrow\infty$ limit corresponds to
$\beta^2\rightarrow \infty$, and in this case $F_{tr}$ as well as
the metric functions $f(r)$ restore the result of EMd with unusual
asymptotic \cite{Shey3}.}

{Furthermore, from the dilaton field (\ref{phi}) we see that as
$r\rightarrow\infty$, the dilaton field does not vanishes, while
in case of asymptotic flat or (A)dS we expect to have $
\lim_{r\longrightarrow \infty} \Phi(r)=0. $ Indeed, by solving the
field equation (\ref{FE2}) we find
\begin{equation}
\Phi(r)=\frac{(n-2)\alpha }{2(\alpha ^{2}+1)}\ln
\left(a+\frac{b}{r}\right),
\end{equation}
however, the system of equation (\ref{FE1})-(\ref{FE3}) will be
fully satisfied provided we choose $a=0$. From the above arguments
we conclude that the asymptotic behaviour of the obtained
solutions is neither flat nor (A)dS.}

\begin{figure}[tbp]
\epsfxsize=7cm \centerline{\epsffile{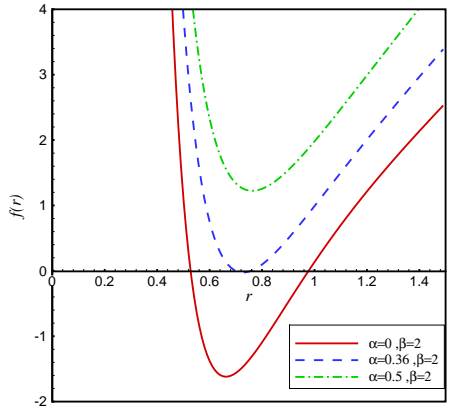}} \caption{ $f(r)$
versus $r$ for $q=2$, $n=6$ and  $m=2.5$.} \label{figure5}
\end{figure}

\begin{figure}[tbp]
\epsfxsize=7cm \centerline{\epsffile{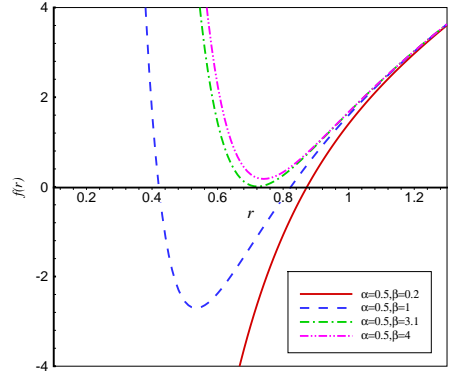}} \caption{$f(r)$
versus $r$ for $q=2$, $m=2.5$ and $n=7$.} \label{figure6}
\end{figure}

It is also worthwhile to note that in case of $\alpha =\sqrt{n-1}$
and $\alpha=1$ the solutions are ill-defined as one can see from
Eq. (\ref{f2}). In order to obtain the location of the horizons of
spacetime, we have to find the roots of $f(r_{+})=0$. However, due
to the complexity of $f(r)$ given in (\ref{f2}), it is not
possible to find the roots of $f(r_{+})=0$, analytically.
Nevertheless, we can plot the function $f(r)$ versus $r$ for
different model parameters as in figures \ref{figure5} and
\ref{figure6}. For simplicity, in these figures, we kept fixed the
other parameters $b=l=1$. Figure \ref{figure5} shows that the
obtained solutions may represent a black hole with two horizons,
an extreme black hole or a naked singularity depending on the
metric parameters. It also shows that for fixed value of the other
parameters, the number of horizons decreases with increasing
$\alpha$. On the other hand from figure \ref{figure6} we see that
for fixed value of $m$, $\alpha$ and $q$, there is a minimum
($\beta_{\rm min}$) and extreme $(\beta_{\rm ext})$ value for the
nonlinear parameter for which we have black hole with a
non-extreme horizon provided $\beta\leq\beta_{\rm min}$, black
hole with two horizons for $\beta_{\rm min}<\beta<\beta_{\rm
ext}$, black hole with an extreme horizon for $\beta=\beta_{\rm
ext}$ and naked singularity for $\beta>\beta_{\rm ext}$. Clearly,
$\beta_{\rm min}$ and $\beta_{\rm ext}$ depend on the other
parameters of the model. {It is worth mentioning that $\beta_{\rm
ext}$ is the value of $\beta$ in which the two horizons meet and
our black hole has only one horizon. In other words, $f(r)=0$ has
a degenerate solution. This correspond to green curve in figure
\ref{figure6}. Besides in this figure, the red curve corresponds
to $\beta\leq\beta_{\rm min}$, the blue curve shows the case
$\beta_{\rm min}<\beta<\beta_{\rm ext}$, and the purple curve
indicates the case with $\beta>\beta_{\rm ext}$. On the other
hand, since $\beta_{\rm ext}$ corresponds to the minimum value of
$f(r)$, thus in this case $f'(r)\mid_{ r=r_{\rm ext}}=0$, which
implies that the surface gravity should be vanished.}
\begin{figure}[tbp]
\epsfxsize=7cm \centerline{\epsffile{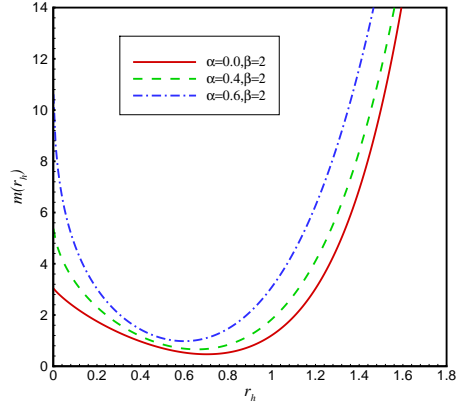}} \caption{The mass
parameter $m$ versus $r_{+}$ for different $\alpha$ and $n=8$ and
$q=1$. } \label{figure7}
\end{figure}
\begin{figure}[tbp]
\epsfxsize=7cm \centerline{\epsffile{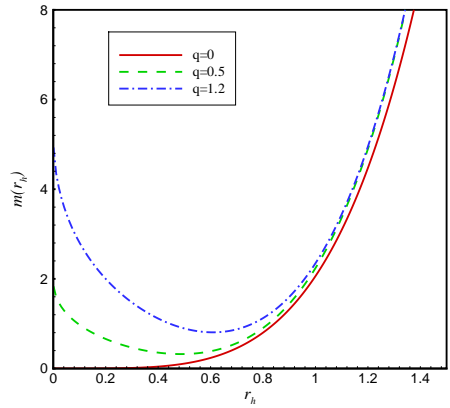}} \caption{The mass
parameter $m$ versus $r_{+}$ for different value of the charge
parameter $q$. Here we have taken $\alpha=0.5$, $n=8$ and
$\beta=1$.} \label{figure8}
\end{figure}

\begin{figure}[tbp]
\epsfxsize=7cm \centerline{\epsffile{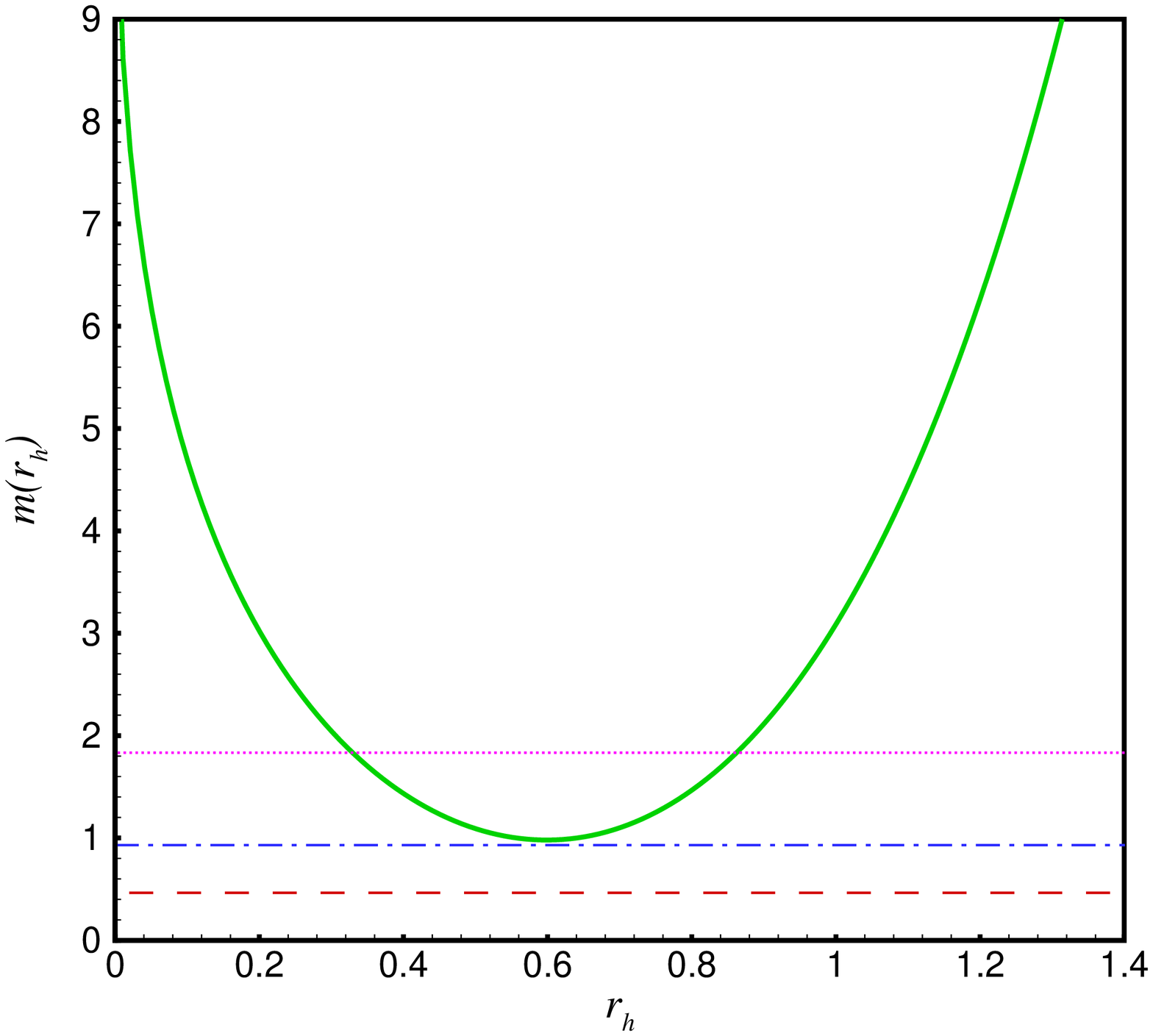}} \caption{The mass
parameter $m$ versus $r_{h}$ for $\alpha=0.6$, $n=8$, $\beta=2$
and $q=1$. $m<m_{\rm ext}$ (red line), $m=m_{\rm ext}$ (blue
line), $m>m_{\rm ext}$ (purple line).} \label{figure8m}
\end{figure}

The nature of the horizons can be further understood if we plot
the mass parameter $m$ as a function of the horizon radius $r_{h}$
for different model parameters. Solving Eq. $f(r_{h})=0$, for the
mass parameter yields
 \begin{eqnarray}\label{mrh}
m(r_{h})&=&-\frac{(n-3)( {\alpha}^{2}+1)^2}{( {\alpha}^{2}+n-3)(
{\alpha}^{2}-1)
}{b}^{-\gamma}{r_{h}}^{2\gamma+n-3-n\gamma/2}+\frac{4(\Lambda+2\beta^2)
\left( {\alpha}^{2}+1 \right)
^{2}{b}^{\gamma}}{(2n-4)(\alpha^{2}-n+1)
}r_{h}^{n-1-{n\gamma}/{2}}\nonumber\\&&+\frac{4 \beta
q{(\alpha^2+1)^2}{b^{(4-n)\gamma/2}}}{n-2}{\left(\frac{\beta^2
b^{(n-2)\gamma}}{q^2} \right)
}^{\frac{1-\gamma}{(n-2)(\gamma-2)}}{L_W^{\frac{\alpha^2-n+1}{2n-4}}(\eta_{h})}
\nonumber
\\ &&\times\Bigg{\{}\frac{L_W^2(\eta_{h})}{\alpha^2+3n-7}\digamma\left(\frac{\alpha^2+3n-7}{2n-4},\frac{\alpha^2+5n-11}{2n-4},\frac{\alpha^2-1}{2n-4}L_W(\eta_{h})\right)
\nonumber\\&&
-\frac{1}{\alpha^2-n+1}\digamma\left(\frac{\alpha^2-n+1}{2n-4},\frac{\alpha^2+n-3}{2n-4},\frac{\alpha^2-1}{2n-4}L_W(\eta_{h})\right)
\Bigg{\}},
\end{eqnarray}
where $\eta_{h}=\eta(r=r_{h})$. Figures \ref{figure7},
\ref{figure8} and \ref{figure8m} show that for fixed value of
other parameters, the value of $m$ determines the number of
horizons. For simplicity in these figure we set $l=b=1$. {From
figure \ref{figure8m} we see that if we solve Eq. $m=\rm const.$
for $r_{h}$, we can distinguish three cases depending on the value
of $m$. For $m>m_{\rm ext}$, there exist two value for
$r_{h}=r_{\pm}$ and thus we have two horizons, for $m=m_{\rm ext}$
the two horizons meet. In this case we encounter an extremal black
hole with zero temperature. As we will show in the next section
the extremal black hole with mass $m_{\rm ext}$ and degenerate
horizon has zero temperature. Besides, for $m<m_{\rm ext}$ there
is no horizon. Furthermore, from figure \ref{figure8} we see that
in the limit $r_{h}\rightarrow0$ we have a nonzero value for the
mass parameter $m$. This is in contrast to the Schwarzschild black
holes in which the mass parameter goes to zero as
$r_{h}\rightarrow0$.   This is due to the effect of the
nonlinearity of the electrodynamic field and in case of $q=0$, the
mass parameter $m$ goes to zero as $r_{h}\rightarrow0$. Physically
$r_{h}\rightarrow 0$ means that the radius of the horizon becomes
very small. Since the horizon radius depends on the solutions
parameters such as $m$, thus we have plotted this behavior in
figure \ref{figure8}. For the simple Schwarzschild black hole
$m=\frac{r_{h}}{2}$ and so as $r_{h}\rightarrow 0$ we have
$m\rightarrow 0$.}
%%%%%%%%%%%%%%%%%%%%%%%%%%%%%%%%%%%%%%%%%%%%%%%%%%%%%%%%%%%%%%%%%%
\section{Thermodynamics of dilaton black holes} \label{Therm}
\begin{figure}[tbp]
\epsfxsize=7cm \centerline{\epsffile{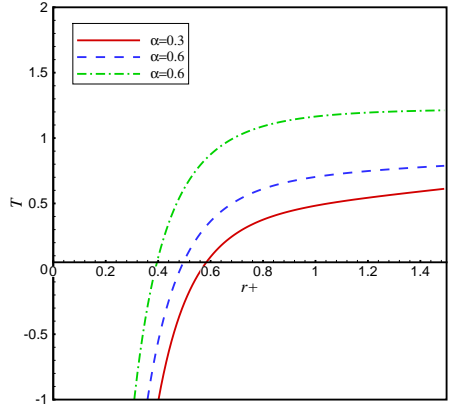}} \caption{ $T$
versus $r_{+}$ for different values of $\alpha$ parameter. Here we
take $l=b=1$, $\beta=2$, $n=5$ and $q=1$. } \label{figure9}
\end{figure}

\begin{figure}[tbp]
\epsfxsize=7cm \centerline{\epsffile{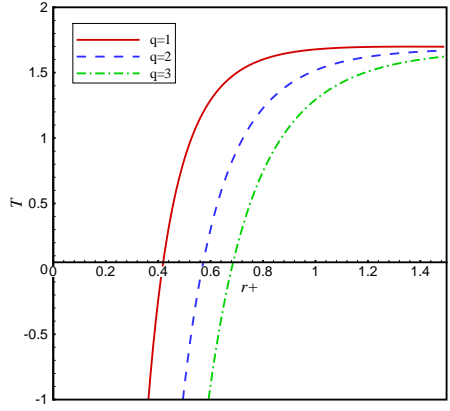}} \caption{$T$
versus $r_{+}$ for different values of charge parameter $q$. Here
we take $l=b=1$, $\alpha=0.8$, $n=6$ and $\beta=2$. }
\label{figure10}
\end{figure}

\begin{figure}[tbp]
\epsfxsize=7cm \centerline{\epsffile{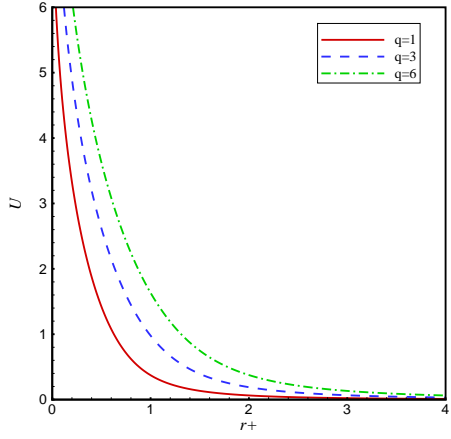}} \caption{The
behaviour of $U(r_{+})$ versus $r_{+}$ for  $\alpha=0.5$,
$\beta=2$ and $n=6$ and different $q$. } \label{figure11}
\end{figure}

\begin{figure}[tbp]
\epsfxsize=7cm \centerline{\epsffile{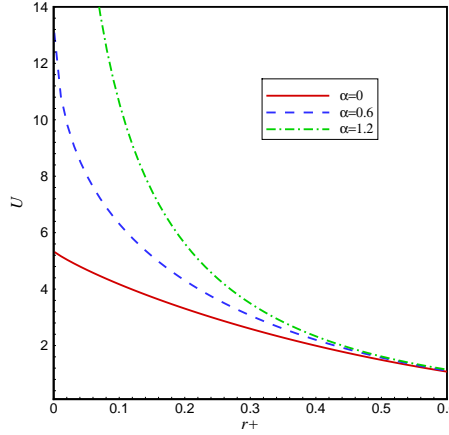}} \caption{
$U(r_{+})$ versus $r_{+}$ for $\beta=2$, $n=8$ and $q=1$ and
different $\alpha$.} \label{figure12}
\end{figure}

{The Hawking temperature of the black hole on the outer horizon
$r_{+}$, may be obtained through the use of the definition of
surface gravity \cite{HendiJHEP}
\begin{eqnarray}\label{Tem1}
T_{+}&=&\frac{\kappa}{2\pi}=\frac{1}{2\pi}\sqrt{-\frac{1}{2}(\nabla_{\mu}\chi
_{\nu}) (\nabla^{\mu}\chi^{\nu})},
\end{eqnarray}
where $\kappa$ is the surface gravity and $\chi=\partial/\partial
t$ is the null killing vector of the horizon. Taking
$\chi^{\nu}=(-1,0,0,...)$, we have $\chi_{\nu}=(f(r_{+}),0,0,...)$
and hence $(\nabla_{\mu}\chi _{\nu})
(\nabla^{\mu}\chi^{\nu})=-\frac{1}{2}\left[f'(r_{+})\right]^2$
which leads to
\begin{eqnarray}\label{kappa}
\kappa=\sqrt{-\frac{1}{2}(\nabla_{\mu}\chi _{\nu})
(\nabla_{\mu}\chi_{\nu})}=\frac{1}{2}
\left(\frac{df(r)}{dr}\right)_{r=r_{+}}.
\end{eqnarray}
Thus, the temperature is obtained as}
\begin{eqnarray}\label{Tem2}
T_{+}&=&\frac{f^{\text{ }^{\prime
}}(r_{+})}{4\pi}=-\frac{(\alpha^2+1)}{4\pi}{r_{+}
^{1-\gamma}}\Bigg{\{}\frac{(n-3) b^{-\gamma}
r_{+}^{2\gamma-2}}{\alpha^2-1}+\frac{2(\Lambda+2\beta^2)b^{\gamma}}{n-2}\nonumber\\&&-\frac{4\beta
q}{n-2}
r_{+}^{{(n-2)(\gamma-2)}/{2}}{b^{{(4-n)\gamma}/{2}}}\left(\frac{1}{\sqrt{L_{W}(\eta_{+})}}-\sqrt{L_{W}(\eta_{+})}\right)\Bigg{\}},
\end{eqnarray}
where $\eta_{+}=\eta(r=r_{+})$ and we have used $f(r_{+})=0$ for
deleting $m$. The behavior of $T$ versus $r_{+}$ is shown in
figures \ref{figure9} and \ref{figure10}. From these figures we
find out that, for large value of $r_{+}$, the temperature tends
to a constant independent of the model parameters, while for small
values of $r_{+}$, the temperature may be negative ($T < 0$). In
this case we encounter a naked singularity. On the other hand, for
an extremal black hole the temperature is zero and the horizon is
degenerate. In this case $r_{\rm ext}$ is the positive root of the
following equation:
\begin{eqnarray}\label{rext}
&&\frac{(n-3) b^{-\gamma} r_{\rm
ext}^{2\gamma-2}}{\alpha^2-1}+\frac{2(\Lambda+2\beta^2)b^{\gamma}}{n-2}\nonumber\\&&-\frac{4\beta
q_{\rm ext}}{n-2} r_{\rm
ext}^{{(n-2)(\gamma-2)}/{2}}{b^{{(4-n)\gamma}/{2}}}\left(\frac{1}{\sqrt{L_{W}(\eta_{\rm
ext})}}-\sqrt{L_{W}(\eta_{\rm ext})}\right)=0.
\end{eqnarray}
where
\begin{equation}
\eta_{\rm ext}\equiv \frac{q_{\rm ext}^{2}r_{\rm
ext}^{(n-2)(\gamma-2)}}{\beta ^{2}b^{(n-2)\gamma}}. \label{etaext}
\end{equation}
From figures \ref{figure9}  and \ref{figure10}, we see that
$r_{\rm ext}$ decreases as $\alpha$ increases, while $r_{\rm ext}$
increases with increasing $q$. Indeed, the metric of Eqs.
(\ref{metric}) and (\ref{f2}) can describe a nonlinear dilaton
black hole with inner and outer event horizons located at $r_{-}$
and $r_{+}$, provided $r>r_{\mathrm{ext}}$, an extreme ENd black
hole in the case of $r=r_{\mathrm{ext}}$, and a naked singularity
if $r<r_{\mathrm{ext}}$. Note that in the limiting case where
$\beta\rightarrow\infty$, expression (\ref{Tem2}) reduces to the
temperature of higher dimensional EMd black holes \cite{Shey3},
\begin{eqnarray}\label{Tem3}
T_{+}&=&-\frac{b^{-\gamma}(\alpha ^2+1)(n-3)}{4
\pi(\alpha^2-1)}r_{+}^{\gamma-1}-\frac{\Lambda (\alpha ^2+1)
b^{\gamma}}{2(n-2)
\pi}r_{+}^{1-\gamma}-\frac{q^{2}b^{-\gamma(n-3)} (\alpha
^2+1)}{2\pi(n-2) }r_{+}^{(\gamma{n}+5-3\gamma-2n)}+O
\left(\frac{1}{\beta^2}\right).
\end{eqnarray}
The entropy of the ENd black hole still obeys the so called area
law of the entropy which states that the entropy of the black hole
is a quarter of the event horizon area \cite{Beck}. This near
universal law applies to almost all kinds of black holes,
including dilaton black holes, in Einstein gravity \cite{hunt}. It
is easy to show
\begin{equation}
{S}=\frac{
b^{(n-2)\gamma/2}r_{+}^{(n-2)(1-\gamma/2)}\omega_{n-2}}{4}.\label{Entropy}
\end{equation}
{The gauge potential $A_{t }$ corresponding to the electromagnetic
field (\ref{FtrE2}) can be obtained through relation
$F_{\mu\nu}=\partial_{\mu}A_{\nu}-\partial_{\nu}A_{\mu}$. Since
our solution is static, thus the gauge potential is only a
function of $r$. From $F_{tr}=\partial_{t}A_{r}-\partial_{r}A_{t}$
with $\partial_{t}A_{r}=0$ we have
\begin{equation}
F_{tr}+\partial_{r}A_{t}(r)=0,
\end{equation}
and hence the gauge potential $A_{t }$ can be derived as
\begin{equation}
A_{t}(r)=-\int{F_{tr} dr}=-\beta b^\gamma \int{ dr r^{-\gamma}
\sqrt{L_W(\eta)}}
\end{equation}
where $\eta$ is defied in (\ref{eta}). Integrating yields
\begin{eqnarray}\label{vectorpot}
A_{t}&=&b^{\gamma}\beta(\alpha^2+1)\left(\frac{\beta^2
b^{(n-2)\gamma}}{q^2}\right)^{\frac{\alpha^2-1}{2n-4}}\left(\frac{\alpha^2-1}{4-2n}\right)^{\frac{\alpha^2+n-3}{4-2n}}\nonumber
\\ && \times \Bigg{\{}
-\frac{1}{2(n-2)}\Gamma\left(\frac{\alpha^2+n-3}{2n-4},\frac{1-\alpha^2}{2n-4}L_{W}(\eta)\right)\nonumber\\&&
+\frac{1}{\alpha^2-1}\left[\Gamma\left(\frac{\alpha^2+3n-7}{2n-4},\frac{1-\alpha^2}{(2n-4)}L_{W}(\eta)\right)-\frac{1}{2}\Gamma\left(\frac{\alpha^2+n-3}{2n-4}\right)\right]
\Bigg{\}}.
\end{eqnarray}
The electric potential $U$, measured at infinity with respect to
the horizon, is defined by \cite{DHSR}
\begin{equation}
U=A_{\mu }\chi ^{\mu }\left| _{r\rightarrow \infty }-A_{\mu }\chi
^{\mu }\right| _{r=r_{+}},  \label{Pot}
\end{equation}
where $\chi=\partial_{t}$ is the null generator of the horizon. It
is a matter of calculation to show that}
\begin{eqnarray}
U&=&b^{\gamma}\beta(\alpha^2+1)\left(\frac{\beta^2
b^{(n-2)\gamma}}{q^2}\right)^{\frac{\alpha^2-1}{2n-4}}\left(\frac{\alpha^2-1}{4-2n}\right)^{\frac{\alpha^2+n-3}{4-2n}}\nonumber
\\ && \times \Bigg{\{}
-\frac{1}{2(n-2)}\Gamma\left(\frac{\alpha^2+n-3}{2n-4},\frac{1-\alpha^2}{2n-4}L_{W}(\eta_{+})\right)\nonumber\\&&
+\frac{1}{\alpha^2-1}\left[\Gamma\left(\frac{\alpha^2+3n-7}{2n-4},\frac{1-\alpha^2}{2n-4}L_{W}(\eta_{+})\right)-\frac{1}{2}\Gamma\left(\frac{\alpha^2+n-3}{2n-4}\right)\right]
\Bigg{\}}. \label{Pot}
\end{eqnarray}
Expanding for large value of $\beta$, we get
\begin{eqnarray}
U=\frac{q
(\alpha^2+1)b^{(4-n)\gamma/2}}{(\alpha^2+n-3)r_{+}^{n-3+2\gamma-n
\gamma/2}} +O\left(\frac{1}{\beta^2}\right).
\end{eqnarray}
We have shown the the behavior of the electric potential $U$ as a
function of horizon radius $r_{+}$ in figures \ref{figure11} and
\ref{figure12} for $b=1$. Due to the nature of the nonlinear
electrodynamics, the electric potential can be finite as
$r_{+}\rightarrow0$, depending on the model parameters, and goes
to zero for large $r_{+}$ independent of the model parameters.
From these figures we find that for fixed value of other
parameter, the divergency of $U$, for small $r_{+}$, increases
with increasing $\alpha$ and $\beta$.
{There are several ways for calculating the mass of the black
holes. For example, for asymptotically AdS solution one can use
the conterterm method inspired by (A)dS/CFT correspondence
\cite{Mal,Haro}. Another way for calculating the mass is through
the use of the substraction method of Brown and York \cite{BY}.
Such a procedure causes the resulting physical quantities to
depend on the choice of reference background. In our case, due to
the presence of the non-trivial dilaton field, the asymptotic
behaviour of the solutions are neither flat nor (A)dS, therefore
we have used the reference background metric and calculate the
mass. According to the substraction method of \cite{BY}, if we
write the metric of $n$-dimensional static  and spherically
symmetric spacetime in the form \cite{CHM}
\begin{equation}\label{metric2}
ds^2=-W^2(r)dt^2 + {dr^2\over V^2(r)}+ r^2d\Omega_{n-2}^2,
\end{equation}
and the matter action contains no derivatives of the metric, then
the quasilocal mass is given by \cite{CHM}
\begin{equation}\label{QLM}
{\cal M} = \frac{n-2}{2}r^{n-3}W(r)\left( V_{0}(r)-V(r)\right).
\end{equation}
Here $V_{0}(r)$ is an arbitrary function which determines the zero
of the energy for a background spacetime and $r$ is the radius of
the spacelike hypersurface boundary. It was argued that the $ADM$
mass $M$ is the ${\cal M}$ determined in (\ref{QLM}) in the limit
$r\rightarrow\infty$ \cite{CHM}. Transforming  metric
(\ref{metric}) in the form (\ref{metric2}), the mass of the
solution is obtained as}
\begin{equation}
{M}=\frac{b^{(n-2)\gamma/2}(n-2) \omega
_{n-2}}{16\pi(\alpha^2+1)}m. \label{Mass}
\end{equation}
Next, we want to check the first law of thermodynamics for
$n$-dimensional ENd black holes. For this purpose, we first obtain
the mass $M$ as a function of extensive quantities $S$ and $Q$.
Combining expressions for the charge, the mass and the entropy
given in Eqs. (\ref{Charge}), (\ref{Mass}) and (\ref{Entropy}),
and using the fact that $f(r_{+})=0$, we obtain a Smarr-type
formula as
\begin{eqnarray}\label{Msmar1}
M(S,Q)&=&-\frac{
b^{-\alpha^2}{(n-2)(n-3)(\alpha^2+1)}{\left(4S\right)}^{\frac{\alpha^2+n-3}{n-2}}
}{16\pi(\alpha^2+n-3)(\alpha^2-1)}
+\frac{(\alpha^2+1)b^{\alpha^2}{(\Lambda+2\beta^2)}}{8\pi
(\alpha^2-n+1)} {\left(4S\right)}^{\frac{-\alpha^2+n-1}{n-2}}
\nonumber\\
&&+\frac{b^\gamma\beta
Q}{(2n-4)}\frac{(\alpha^2+1)^3}{(\alpha^2-1)^2}{\left(\frac{\beta^2
b^{(n-2)\gamma}}{16\pi^2
Q^2}\right)}^{\frac{\alpha^2-1}{2n-4}}\left(\frac{1-\alpha^2}{2n-4}\right)^{\frac{\alpha^2+1-n}{4-2n}}\nonumber
\\ && \times \Bigg{\{} -(n-2)^2(\gamma-2)^2\left[\Gamma
\left(\frac{\alpha^2+3n-7}{2n-4},\frac{1-\alpha^2}{2n-4}L_{W}\left(\zeta\right)\right)-\Gamma\left(\frac{\alpha^2+3n-7}{2n-4}\right)\right]\nonumber \\
&&+(\gamma-1)^2\left[\Gamma
\left(\frac{\alpha^2-n+1}{2n-4},\frac{1-\alpha^2}{2n-4}L_{W}\left(\zeta\right)\right)
-\Gamma\left({\frac{\alpha^2-n+1}{2n-4}}\right)\right]\Bigg{\}},
 \end{eqnarray}
where $\zeta=\frac{\pi^2Q^2}{S^2\beta^2}$. If we expand $M(S,Q)$
for large $\beta$, we arrive at
\begin{eqnarray}
 M(S,Q)&=&-\frac{ b^{-\alpha^2}(n-2)(n-3)(\alpha^2+1){\left(4S\right)}^{\frac{\alpha^2+n-3}{n-2}}
}{16\pi(\alpha^2+n-3)(\alpha^2-1)}
+\frac{\Lambda(\alpha^2+1)b^{\alpha^2}}{8\pi
(\alpha^2-n+1)}{\left(4S\right)}^{\frac{-\alpha^2+n-1}{n-2}}\nonumber
\\&& +\frac{2\pi (\alpha^2+1)Q^2
b^{\alpha^2}}{\alpha^2+n-3}(4
S)^{\frac{\alpha^2+n-3}{2-n}}+O\left(\frac{1}{\beta^2}\right),
\label{Msmar2}
 \end{eqnarray}
which is exactly the Smarr-type formula obtained for EMd black in
the limit $\beta\rightarrow\infty$ \cite{Shey3}. Now, if we
consider $S$ and $Q$ as a complete set of extensive parameters for
the mass $M(S,Q)$, we can define the intensive parameters
conjugate to $S$ and $Q$ as
\begin{equation}
T=\left( \frac{\partial M}{\partial S}\right) _{Q},\ \ U=\left( \frac{\partial M%
}{\partial Q}\right) _{S}.  \label{Dsmar}
\end{equation}
Numerical calculations show that the intensive quantities
calculated by Eq. (\ref{Dsmar}) coincide with Eqs. (\ref{Tem2})
and (\ref{Pot}). Thus, these thermodynamic quantities satisfy the
first law of black hole thermodynamics
\begin{equation}
dM = TdS+Ud{Q}.
\end{equation}
The satisfaction of the first law of thermodynamics for the
obtained conserved and thermodynamic quantities, together with the
fact that these quantities in two limiting cases, namely in the
absence of the dilaton field ($\alpha=0=\gamma$), and for large
values of the nonlinear parameter ($\beta\rightarrow\infty$),
reduce to the known results in the literature \cite{Hendi2,Shey3},
indicate that the conserved and thermodynamic quantities obtained
in this paper are correct and in agreement with other method such
as Euclidean action method \cite{Mis}.
\section{conclusion and discussion}
In this paper, we generalized the study on the EN electrodynamics
by taking into account the dilaton scalar field in the action. We
first proposed the suitable Lagrangian for EN electrodynamics
coupled to the dilaton field and in the presence of two
Liouville-type dilaton potential for the dilaton field in all
higher dimensions. By varying the action we obtained the field
equations of $n$-dimensional EN electrodynamics coupled to dilaton
field in Einstein gravity. Then, we constructed a new class of
higher dimensional static and spherically symmetric black hole
solutions of this theory. When $\beta\rightarrow\infty$, our
solutions reduce to higher dimensional EMd black hole solutions
\cite{Shey3}, while in the absence of the dilaton field,
($\alpha=0=\gamma$), they restore charged black holes coupled to
EN electrodynamics. Although the behavior of the electric field
near the origin depends on the model parameters, however for large
$r$ the asymptotic behavior of electric field are exactly the same
as linear Maxwell field. Interestingly enough, we found that the
electric field of ENd black hole is finite near the origin and
diverges exactly at $r=0$ depending on the model parameters,
however its divergency is much weaker than Maxwell field. Besides,
in the absence of the dilaton field ($\alpha=0$), the electric
field has a finite value near $r=0$, while as soon as the dilaton
field is taken into account ($\alpha>0$), the electric field
diverges as $r\rightarrow 0$. This implies that the presence of
the dilaton field changes the behaviour of the electric field near
the origin where $r=0$.

We also found that the dilaton field changes the asymptotic
behavior of the solutions to be neither flat nor (A)dS. This is
consistent with the fact that no dilaton dS or AdS black hole
solution exists with the presence of only one or two
Liouville-type dilaton potential \cite{MW}. The obtained solutions
can represent black holes with inner and outer horizons, an
extreme black hole or naked singularity depending on the model
parameters. For fixed value of $m$, $\alpha$ and $q$, we found
that there is a minimum (extreme) value for nonlinear parameter
$\beta_{\rm min}$ $(\beta_{\rm ext})$, for which we have black
hole with a non-extreme horizon provided $\beta\leq\beta_{\rm
min}$, black hole with two horizons for $\beta_{\rm
min}<\beta<\beta_{\rm ext}$, black hole with an extreme horizon
for $\beta=\beta_{\rm ext}$ and naked singularity for
$\beta>\beta_{\rm ext}$. We computed conserved and thermodynamic
quantities and obtained Smarr-type formula,$M(S,Q)$. We checked
that the conserved and thermodynamic quantities obtained for
higher dimensional  dilaton black holes satisfy the first law of
black holes thermodynamics on the horizon.

Let us emphasize that the higher dimensional dilaton black holes
coupled to EN electrodynamics constructed in this paper are
static. Thus, it would be interesting to derive $n$-dimensional
rotating black hole/brane solutions of these field equations. The
study can also be applied for other type of nonlinear
electrodynamics such as logarithmic nonlinear electrodynamics with
suitable lagrangian in the presence of dilaton field. It is also
intersecting to study stability of the solutions. These issues are
now under investigations and the results will be appeared
elsewhere.

%%%%%%%%%%%%%%%%%%%%%%%%%%%%%%%%%%%%%%%%%%%%%%%%%%%%%%%%%%%%%%%%%%%%%%%%%%%%%%%%%
\acknowledgments{A. Sheykhi thanks Shiraz University Research
Council. This work has been supported financially by Center for
Excellence in Astronomy and Astrophysics of IRAN (CEAAI),
Maragheh, IRAN.}
%%%%%%%%%%%%%%%%%%%%%%%%%%%%%%%%%%%%%%%%%%%%%%%%%%%%%%%%%%%%%%%%%%

\end{document}